# Giant Field Enhancement in Longitudinal Epsilon Near Zero Films

Mohammad Kamandi,[1] Caner Guclu,[1] Ting Shan Luk,[2,3] George T. Wang,[2], and Filippo Capolino[1*]

[1]*Department of Electrical Engineering and Computer Science, University of California, Irvine, California 92697, USA*
[2]*Sandia National Laboratories, Albuquerque, New Mexico 87185, USA*
[3]*Center for Integrated Nanotechnologies, Sandia National Laboratories, Albuquerque, NM 87185, USA*
*f.capolino@uci.edu*

We report that a longitudinal epsilon-near-zero (LENZ) film leads to giant field enhancement and strong radiation emission of sources in it and that these features are superior to what found in previous studies related to isotropic ENZ. LENZ films are uniaxially anisotropic films where relative permittivity along the normal direction to the film is much smaller than unity, while the permittivity in the transverse plane of the film is not vanishing. It has been shown previously that realistic isotropic ENZ films do not provide large field enhancement due to material losses, however, we show the loss effects can be overcome using LENZ films. We also prove that in comparison to the (isotropic) ENZ case, the LENZ film's field enhancement is not only remarkably larger but it also occurs for a wider range of angles of incidence. Importantly, the field enhancement near the interface of the LENZ film is almost independent of the thickness unlike for the isotropic ENZ case where extremely small thickness is required. We show that for a LENZ structure consisting of a multilayer of dysprosium-doped cadmium oxide and silicon accounting for realistic losses, field intensity enhancement of 30 is obtained which is almost 10 times larger than that obtained with realistic ENZ materials.

Materials with extremely small permittivity, namely epsilon near zero (ENZ) materials have been at the focus of attention due to their natural existence in optical frequencies and their unprecedented properties. Realization of ENZ behavior has been achieved using multilayer stack of metal and dielectric [1], 3-D periodic array of dielectric-core metallic-shell nanospheres with fluorescent dyes in the core of each nanoparticle for the loss-compensation [2] or employing metal-coated waveguides at their cut-off frequency [3]. Owing to their extremely large velocity of phase propagation, such materials enable linear applications such as tailoring radiation emission [4–7], energy squeezing and supercoupling [8]. On the other hand ENZ materials can be utilized to achieve huge field enhancement. In [9] the field intensity enhancement (FIE) of a isotropic ENZ semi-infinite medium and a isotropic ENZ slab under TM (transverse magnetic) plane wave incidence are theoretically investigated. Exploiting this ability, optical nonlinearities such as second or third harmonic generation [10–15] and Kerr nonlinearities [16] have been enhanced significantly.

In the present letter we establish that under TM wave incidence a uniaxially anisotropic epsilon near zero film exhibits remarkably stronger FIE than isotropic epsilon near zero. Hereafter, we will use IENZ for isotropic epsilon near zero studied in [9] for comparison. The film whose surfaces are normal to the $z$ axis, shown in Fig. 1, is marked by the subscript '2', and modeled via a relative permittivity tensor $\underline{\varepsilon}_2 = \varepsilon_t \left( \hat{\mathbf{x}}\hat{\mathbf{x}} + \hat{\mathbf{y}}\hat{\mathbf{y}} \right) + \varepsilon_z \hat{\mathbf{z}}\hat{\mathbf{z}}$. Particularly we show that the specific type of anisotropy useful for super-field enhancement occurs when the $zz$ entry of the permittivity tensor is near zero, which in the following we call it longitudinal epsilon near zero (LENZ) condition. Most interestingly, we show that FIE in LENZ films occurs for a very wide range of angles of incidence and is almost independent of the film thickness unlike IENZ films where such features occur for a fix angle and extremely thin films [9]. To the best of our knowledge, significant field enhancement can't be achieved using realistic ENZ materials due to inherent material losses. Remarkably, in this paper, we introduce a LENZ structure that provides large field enhancement despite having realistic loss which paves the way for a wide range of applications associated to second harmonic generation and enhanced field emission. Indeed, through reciprocity, we demonstrate that a $z$-polarized dipole located in the LENZ film has stronger far field radiation compared to the IENZ case.

The geometry of the investigated problem is depicted in Fig. 1. We first investigate the FIE in a film with thickness $d$ under a TM plane wave as in Fig. 1(a), and then we investigate the radiative emission enhancement of a point dipole inside a LENZ film as illustrated in Fig. 1(b).

The electric field vector of the incident TM wave is in the $x - z$ plane, i.e., $\mathbf{E}_1^i = E_1^i \left( \cos\theta \hat{\mathbf{x}} + \sin\theta \hat{\mathbf{z}} \right) e^{i\mathbf{k}_1 \cdot \mathbf{r}}$ in which $\mathbf{k}_1$ is the wavevector of the impinging TM wave where $k_1 = |\mathbf{k}_1| = \omega\sqrt{\mu_0 \varepsilon_0 \varepsilon_1}$ is the wavenumber in medium 1. A monochromatic, time harmonic convention $e^{-i\omega t}$ is implicitly assumed. The transverse (to the $z$ axis) wavenumber is $k_t$ whereas the longitudinal wavenumber outside the film is $k_{z1} = \sqrt{k_1^2 - k_t^2}$. In the LENZ film the entries of the relative permittivity tensor are $\varepsilon_t = \varepsilon_t' + i\varepsilon_t''$ and $\varepsilon_z = \varepsilon_z' + i\varepsilon_z''$. We will use $k_{z2} = \sqrt{\varepsilon_t k_0^2 - (\varepsilon_t / \varepsilon_z) k_t^2}$ to denote the longitudinal wavenumber in the film. Owing to the continuity of the normal displacement field component at $z = d/2$,

$$\varepsilon_1 E_{z1}\big|_{z=(d/2)^+} = \varepsilon_z E_{z2}\big|_{z=(d/2)^-} , \quad (1)$$

in which $E_{z1}$ and $E_{z2}$ are the longitudinal components of the total electric field in media 1 and 2 respectively. By replacing the value of $E_{z1}$ in the abovementioned equation one obtains [9]

$$\varepsilon_1 E_1^i (1-\Gamma) \sin\theta = \varepsilon_z E_{z2} , \quad (2)$$

in which $\Gamma$ is the plane wave reflection coefficient at $z = d/2$, seen from the upper interface and is given by

$$\Gamma(d,\theta) = \frac{-i\left(k_{z2}^2 - k_{z1}^2 \hat{\varepsilon}_t^2\right) s_h}{2 k_{z1} k_{z2} \hat{\varepsilon}_t c_h - i\left(k_{z2}^2 + k_{z1}^2 \hat{\varepsilon}_t^2\right) s_h} , \quad (3)$$

with $s_h = \sin(k_{z2}d)$, $c_h = \cos(k_{z2}d)$ and $\hat{\varepsilon}_t = \varepsilon_t / \varepsilon_1$. Assuming $\hat{\varepsilon}_z = \varepsilon_z / \varepsilon_1$, it is convenient to define the local $z$-polarized field intensity enhancement at $z = (d/2)^-$ as

$$\text{FIE} = \left|\frac{E_{z2}}{E_1}\right|^2 = \left|\frac{(1-\Gamma)\sin\theta}{\hat{\varepsilon}_z}\right|^2 , \quad (4)$$

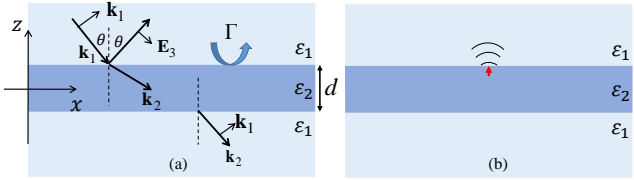

FIG. 1. Schematic of longitudinal epsilon near zero film (a) under TM-plane wave incidence and (b) with dipole located below the interface.

which is the ratio of the electric field in the longitudinal direction in the film to the incident electric field amplitude at the same place in the absence of the film. In the following, unless stated otherwise, FIE is always calculated just below the top surface of the film at $z = (d/2)^-$. The field intensity enhancement depends strongly on the choice of $\varepsilon_z$, i.e., by choosing $\varepsilon_z$ close to zero FIE gets large. FIE is also strongly dependent upon the reflection coefficient $\Gamma$, which in general is complex, and if it gets close to unity then FIE vanishes. We will compare the LENZ and IENZ cases for their field enhancement and radiation enhancement capabilities using examples and analytical calculations.

As an example, in Fig. 2, we consider a film with thickness $d = \lambda/3$, with $\lambda = 2\pi/k_1$, made of LENZ material surrounded by vacuum i.e. $\varepsilon_1 = 1$. In Fig. 2(a) we assume that the film has longitudinal permittivity of $\varepsilon_z = 0.001 + i0.001$ and transverse permittivity of $\varepsilon_t = \varepsilon_t' + i0.001$ at wavelength $\lambda$. We report the FIE at $z = (d/2)^-$, i.e. just below the top surface of the film, versus the real part of the transverse permittivity of the film $\varepsilon_t'$ and the angle of incidence of the impinging TM-polarized wave. The IENZ case, as a subset of LENZ cases reported in Fig. 2 (a), is marked with white dashed line where $\varepsilon_t' = \varepsilon_z'$. Notably, we observe that FIE is the lowest for the IENZ case compared to LENZ cases with larger $\varepsilon_t'$. As the anisotropy of the film becomes starker, the FIE increases significantly, in other words, it is better not to have a vanishing $\varepsilon_t'$. Importantly, the plot shows that LENZ leads to not only larger FIE, but also to a wider angular span of large FIE, contrarily to the IENZ case that provides large FIE only on a very limited angular range [9]. We now exactly show the reason of the physical behavior that differentiates the LENZ from the IENZ: assuming $\varepsilon_1 = 1$ we substitute $\Gamma$ from (3) in (4):

$$\text{FIE} = \left|\frac{2k_{z1}\varepsilon_t \left(k_{z2}c_h - ik_{z1}\varepsilon_t s_h\right)}{\varepsilon_z\left[2k_{z1}k_{z2}\varepsilon_t c_h - i\left(k_{z2}^2 + k_{z1}^2 \varepsilon_t^2\right) s_h\right]}\sin\theta\right|^2 . \quad (5)$$

From this equation one may observe that for isotropic film with permittivity $\varepsilon_2 \to 0$ and $\theta \neq 0$ equation (5) is rewritten as

$$\text{FIE}_{\text{IENZ}} = \left|\frac{2\cos\theta\cos(-ik_1 d \sin\theta)}{\sin(-ik_1 d \sin\theta)}\right|^2 , \quad (6)$$

which is a finite (i.e., not large) value unless $\theta$ or $d$ tends to zero. Note that for an assigned arbitrary $\theta$, $\text{FIE}_{\text{IENZ}}$ does not tend to infinity even if we assume that $\varepsilon_2 \to 0$. It is worth mentioning that $\text{FIE}_{\text{IENZ}}$ in thin ENZ films (i.e., when $d \to 0$) is inversely proportional to the thickness $d$. This is seen by simplifying equation (6) for $d \to 0$ and using $\cos(-ik_1 d \sin\theta) \approx 1$ and $\sin(-ik_1 d \sin\theta) \approx -ik_1 d \sin\theta$, leading to

$$\text{FIE}_{\text{IENZ}} \approx \left|\frac{2\cos\theta}{-ik_1 d \sin\theta}\right|^2 . \quad (7)$$

Instead, for the LENZ case, assuming near zero values for $\varepsilon_z$ and angles such that $\varepsilon_z \ll \sin^2\theta$ (because the proper limit is for $(\varepsilon_z / \sin^2\theta) \to 0$), by simplifying the numerator and denominator of (5), considering finite values of $\varepsilon_t$ and $d$, we obtain

$$\text{FIE}_{\text{LENZ}} \approx \left| \frac{2\sqrt{\varepsilon_t}}{\sqrt{\varepsilon_z}} \cos\theta \right|^2 \approx 4 \left| \frac{\varepsilon_t}{\varepsilon_z} \right| \cos^2\theta . \qquad (8)$$

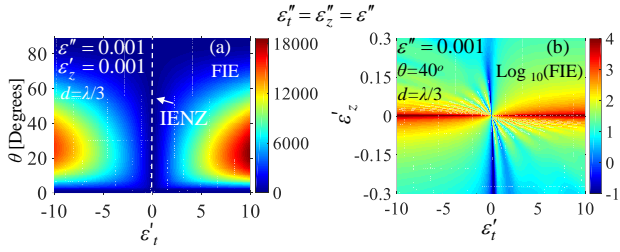

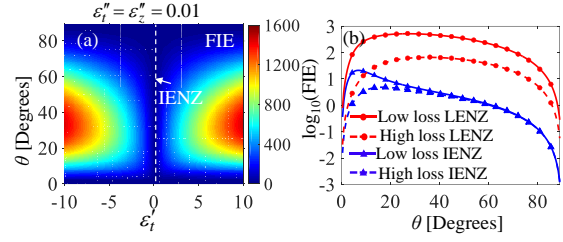

FIG. 2. (a) FIE in LENZ film at $z=(d/2)^-$ in the geometry of Fig. 1 with $d=\lambda/3$ and $\varepsilon_z = 0.001+i0.001$ and $\varepsilon_t = \varepsilon_t' + i0.001$ as a function of $\varepsilon_t'$ and $\theta$. (b) FIE in logarithmic scale versus $\varepsilon_t'$ and $\varepsilon_z'$.

Here the denominator goes to zero as $\varepsilon_z \to 0$ which causes the FIE to tend towards infinity for the LENZ case. Note that to obtain giant FIE is not necessary to illuminate with small incidence angle $\theta$, whereas in the IENZ case only for small $\theta$, one can get giant FIE. The results of FIE are shown in Fig. 2 (b) where FIE is reported in logarithmic scale versus real part of the transverse and longitudinal permittivities for slab with $d=\lambda/3$ and $\varepsilon_t'' = \varepsilon_z'' = 0.001$ under $\theta = 40^o$ incidence angle. Note that as $\varepsilon_z$ tends to zero the FIE value increases. Most importantly, as the film becomes more anisotropic (larger $|\varepsilon_t'|$) FIE increases as well, as was explained with (8). To trace the physical origin of this point, we observe that for a constant near zero value of $\varepsilon_z$ and constant incident angle ($\theta = 40^o$ in this example) FIE only depends on $|1-\Gamma|$. Therefore, the behavior of FIE is a signature of the reflection coefficient $\Gamma$ behavior which tends to one for IENZ unless when $d$ tends to zero. However, for LENZ case, $\Gamma$ does not tend to one, paving the way for obtaining large FIE over a wide range of angles.

One of the most important factors in determining the FIE in LENZ and IENZ films is the loss represented by the imaginary part of the permittivity. Note that to evaluate (6) we have assumed that $\varepsilon_2 \to 0$, however in practical cases one can only choose $\varepsilon_2' \to 0$ and hence $\varepsilon_2 = i\varepsilon_2''$ is the minimum value, that can't be arbitrarily small because of losses, further limiting the FIE growth. For LENZ, ideally FIE tends to infinity when $\varepsilon_z \to 0$, and having large $\varepsilon_t$ is even more favorable for obtaining large FIE, an important aspect not shown in the literature. The presence of losses implies that one can only choose $\varepsilon_z' \to 0$, hence $\varepsilon_z = i\varepsilon_z''$ can't be arbitrarily small, but from (8) a large FIE in LENZ is still obtained when choosing $|\varepsilon_t/\varepsilon_z''|$ to be large, indicating that loss effects are overcome in LENZ. This is a striking result showing that limitations are imposed only by having composite materials with large $\varepsilon_t$. To investigate the effect of the loss, in Fig. 3(a) we reproduce the same set of cases as in Fig. 2(a) but with higher film loss modeled by $\varepsilon_t'' = \varepsilon_z'' = 0.01$ reporting that FIE decreased drastically due to the loss. However LENZ still yields higher FIE compared to IENZ (marked with dashed white line). To better appreciate FIE superiority of LENZ over IENZ in a wide angular range both in low and high loss cases, in Fig. 3(b) FIE is plotted versus incident angle for IENZ with $\varepsilon_2' = 0.001$ and LENZ with $\varepsilon_t' = 2.5$ and $\varepsilon_z' = 0.001$. For the high loss cases we assume $\varepsilon_2'' = 0.05$ for IENZ and $\varepsilon_t'' = \varepsilon_z'' = 0.05$ for LENZ; for the low loss cases we have $\varepsilon_2'' = 0.01$ for IENZ and $\varepsilon_t'' = \varepsilon_z'' = 0.01$ for LENZ. The outstanding performance of LENZ is demonstrated in this figure by noting that high loss LENZ provides much higher FIE even than low loss IENZ for angles of incidence $\theta > 10^o$. With similar imaginary part of permittivity, the FIE of LENZ is two orders of magnitude higher than that for IENZ for a very wide range of angles of incidence. Moreover, the angular range at which FIE occurs is much wider in the LENZ case than in the IENZ case. Using angular full width at half maximum (FWHM) of FIE defined as the range of angles in which FIE is higher than the half of its maximum value, the angular FWHM of FIE in the low-loss LENZ case is at least 45° whereas for the low-loss isotropic case is less than 12°.

FIG. 3. (a) FIE in LENZ film at $z=(d/2)^-$ as in Fig. 2(a) with higher loss $\varepsilon_t'' = \varepsilon_z'' = 0.01$. (b) Comparison between IENZ and LENZ for different losses.

Another important quality of FIE in LENZ is its high value over a range of z-locations within the film. This is reported in Fig. 4(a) as a function of $z$ and $d$, both normalized to wavelength, for a specific case of $\theta = 40^o$, $\varepsilon_z = 0.001+i0.035$ and $\varepsilon_t = 2.5+i0.035$. The FIE is maximum at the interface between the film and air and decreases by getting deeper into the film. For small thicknesses, when $d<0.1\lambda$, the FIE has a more uniform distribution inside the film, and the FIE is at similar levels as in thicker films, so thickness is not important to have large FIE near the interface, contrarily to the IENZ case [9]. In Fig. 4(b) the dependence of FIE on the film thickness ($d/\lambda = 1$, $d/\lambda = 0.1$ and $d/\lambda = 0.01$) is shown as a function of $\varepsilon'_t$ assuming incidence at $\theta = 40^o$, $\varepsilon_z = 0.001 + i0.035$ and $\varepsilon''_t = 0.035$. The exceptional property of a LENZ film with large $|\varepsilon_t|$ to significantly enhance the field independent of its thickness is clearly shown, in contrast to IENZ films where FIE is large only for extremely small thickness. This property is simply understood from equation (8) in which numerator increases with $\varepsilon_t$. Importantly, from Fig. 4 (b) we observe also that very thin films (e.g., $d/\lambda = 0.01$) can provide high FIE almost independently of $|\varepsilon_t|$.

We now discuss the results of a realistic LENZ case obtained with a multilayer structure and compare it to IENZ. The multilayer structure providing LENZ performance when homogenized is made of 10 alternate layers, of equal thickness $\lambda/60$, of Dysprosium-doped cadmium oxide (CdO:Dy) [17] with carrier density $n = 3.7 \times 10^{20}$ cm$^{-3}$, and silicon with permittivity taken from [18]. Using effective medium approximation (EMA) [19], LENZ condition for the homogenized structure occurs at $\lambda_0 = 1867.9$ nm for which $\varepsilon_z = i0.26$ and $\varepsilon_t = 5.98 + i0.065$. In Fig. 5 we calculate the FIE of this structure just below the top surface of the top layer (CdO:Dy) via transfer-matrix method (TMM) and compare it with FIE of bulk CdO:Dy at its ENZ wavelength of $\lambda_0 = 1866.7$ nm with $\varepsilon_2 = i0.13$ (note that $\varepsilon''_z = 2\varepsilon''_2$). Also, to better appreciate the remarkable effect of LENZ, we have provided the FIE of indium tin oxide (ITO) at its ENZ frequency [13]. Films have thicknesses $d = \lambda/3$, where the wavelength is the one at the respective LENZ and ENZ conditions. As it can be seen, not only FIE for the multilayer (LENZ) is higher for all angles of incidence than bulk CdO:Dy but also the maximum of the former is 12 folds the maximum of the latter. The FIE of ITO is even less than the FIE of CdO:Dy due to its higher loss indicating how FIE in ENZ is in general not practical unless you can use $\varepsilon'_t$ as in LENZ to overcome the loss.

Giant z-polarized E-field enhancement inside the film for a wide range of angles of incidence in LENZ also implies, via the reciprocity theorem that a z-polarized dipole located at the E-field hotspot in a LENZ film radiates *very strong far-fields over a wide angular region*. Hence, we show next the capability of LENZ films to enhance a dipole radiation emission. This is described by resorting to the key parameter [20]

$$\text{REE} = P_{\text{rad}} / P_{\text{fs}}, \qquad (9)$$

where REE is the radiative emission enhancement, $P_{\text{rad}}$ is the power radiated in both top and bottom vacuum half-spaces by an impressed dipole located inside the film and $P_{\text{fs}}$ is the total power emitted by the same dipole in free space. $P_{\text{rad}}$ does not account for all the power emitted by that dipole which is also dissipated as loss in the LENZ film. In Fig. 6, REE of a z-polarized dipole inside the LENZ film with thickness $d = \lambda/3$ at an infinitesimal distance from the top surface is plotted versus $\varepsilon'_t$ and $\varepsilon'_z$ for (a) lossless case and (b) when $\varepsilon''_t = \varepsilon''_z = 0.01$. We observe that regardless of the sign of $\varepsilon'_z$, as long as it is small, REE is large. Moreover REE increases as $|\varepsilon'_t|$ increases. In the lossless case, the REE is maximized when $\varepsilon'_t < 0$ and $\varepsilon'_z > 0$ or when $\varepsilon'_t > 0$ and $\varepsilon'_z < 0$, however when losses are introduced, this behavior is less pronounced.

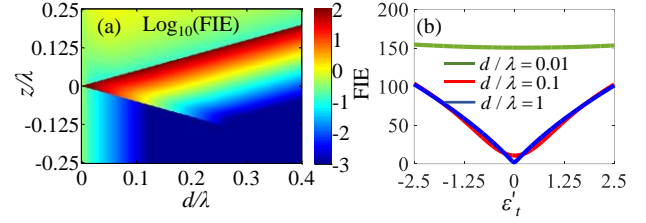

FIG. 4. FIE in LENZ for $\theta = 40^o$, $\varepsilon''_t = \varepsilon''_z = 0.035$ and $\varepsilon'_z = 0.001$ (a) in the film profile for $\varepsilon'_t = 2.5$ (b) as a function of $\varepsilon'_t$ for various thicknesses.

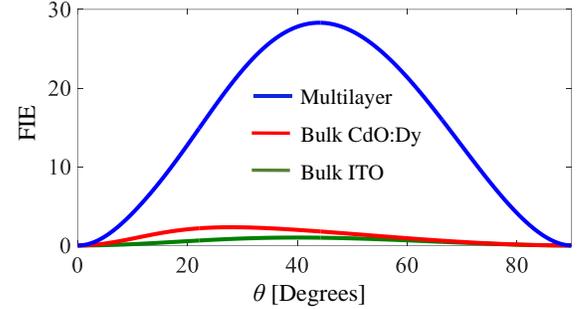

FIG. 5. FIE versus angle of incidence for the realistic LENZ (multilayer) and two IENZ cases: bulk CdO:Dy and bulk ITO.

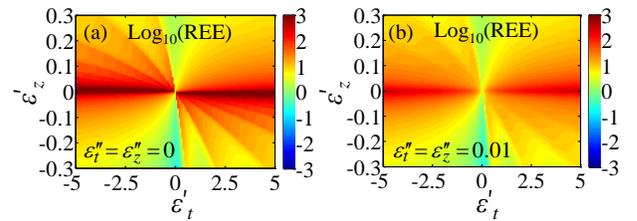

FIG. 6. Radiative emission enhancement, versus $\varepsilon'_t$ and $\varepsilon'_z$ (a) for the lossless case (the color legend is saturated for values more than 1000) and (b) for a lossy case.

In conclusion, we demonstrated the unique ability of LENZ films to generate electric field enhancement and why it is superior to what can be obtained with IENZ. We showed that for the same level of loss, LENZ gives much higher FIE than IENZ and also occurs for a wider range of angles of incidence compared to the IENZ. Furthermore, FIE is almost independent of the thickness of the film unlike the IENZ case where the film has to be extremely thin. Remarkably, losses plays a major role in practical IENZ cases for generating FIE but loss effects is instead overcome in LENZ by increasing $|\varepsilon_t|$. Finally, radiative emission in LENZ is higher than in IENZ films and it occurs over a wide angular region with possible applications also in light generation [21].

Acknowledgments. This work was performed, in part, at the Center for Integrated Nanotechnologies, an Office of Science User Facility operated for the U.S. Department of Energy (DOE) Office of Science. This work was also partly supported by the National Science Foundation, NSF-SNM-1449397.